\documentclass[showpacs,preprintnumbers,amsmath,amssymb,prb,twocolumn,floatfix,amssymb]{revtex4}
\usepackage[dvips]{graphicx,color}
\usepackage{dcolumn}
\usepackage{amsmath,amssymb,bbold,bm}

\begin{document}
\title{Noninvasive Probes of Charge Fractionalization in Quantum Spin-Hall Insulators}
\author{Ion Garate$^1$ and Karyn Le Hur$^{1,2}$}
\affiliation{$^1$ Department of Physics, Yale University, New Haven, CT 06520, USA}
\affiliation{$^2$ Center for Theoretical Physics, Ecole Polytechnique, CNRS,  91128 Palaiseau Cedex, France.}
\date{\today}

\begin{abstract}
When an electron with well-defined momentum tunnels into a nonchiral Luttinger liquid, it breaks up into two separate wave packets that carry fractional charges and move in opposite directions. 
A direct observation of this phenomenon has proven elusive, mainly due to single-particle and plasmon backscattering caused by measurement probes.
This paper theoretically introduces two topological insulator devices that are naturally suited for detecting fractional charges and their velocities directly and in a noninvasive fashion.
\end{abstract}
\maketitle

\section{Introduction}
One dimensional (1D) electron systems, known as Luttinger liquids (LLs), have been long predicted to exhibit a plethora of phenomena that are caused by electron-electron interactions and possess no analogues in higher dimensional liquids.~\cite{deshpande2010} 
One of those predictions is ``charge fractionalization'',~\cite{pham2000} whereby a charge $q$ injected unidirectionally in the middle of a LL breaks apart into two counterpropagating pulses.
These pulses carry definite charge fractions $q (1+g)/2$ and $q (1-g)/2$, where $g$ is the Luttinger parameter that quantifies the strength of electron correlations ($g=1$ for free electrons, $g<1$ for repulsive interactions).

Although the detection of said charge fractions would constitute a milestone toward the characterization of LLs, it remains a challenging endeavor. 
Most experimental difficulties originate from the fact that fractional charges reside inside the LL ($g<1$), whereas measurement probes that couple to the LL are higher-dimensional Fermi liquids ($g=1$).
As a result, fractional charges are partially reflected and hence degraded at the interface between the LL and the detection probe.
Charge reflection occurs in two forms: (i) single-particle backscattering (when the probe-LL interface is not smooth or when there are sharp impurities in the LL) and (ii) wave (or ``plasmon'') backscattering. The latter is present even in atomically smooth LL-probe interfaces, because of a mismatch in $g$ (or equivalently in the plasmon index-of-refraction).~\cite{safi1995} 
In recent years, various proposals have been put forward in order to bypass the aforementioned difficulties: 
dephasing of quantum interference in 1D rings,~\cite{lehur2005} d.c. conductance in three-terminal nanowire geometries,~\cite{steinberg2008} and high-frequency shot noise in unidirectional (chiral) LLs of integer quantum Hall systems,~\cite{berg2009} are all believed to display fingerprints of charge fractionalization.
At any rate, a well-controlled charge measurement that would provide a smoking gun for fractional charges in nonchiral LLs is still nonexistent.

In this paper we intend to help overcome such challenge by designing two two-dimensional (2D) topological insulator (TI) circuits with capacitive coupling to noninvasive charge sensors.
2D TIs, dubbed quantum spin-Hall insulators (QSHI), are insulating in the bulk but endowed with topologically protected conducting edge states.~\cite{hasan2010}
In presence of electron interactions, these edge states behave as helical (nonchiral) LLs,~\cite{wu2006} whose right- and left-moving excitations are spin-polarized along opposite directions.
The main message from the present study is that helical LLs constitute favorable platforms to measure fractional charges and their propagation speeds directly and without distortion.

\section{Proposed Experimental Setups}

\begin{figure}[t]
\begin{center}
\includegraphics[scale=0.42]{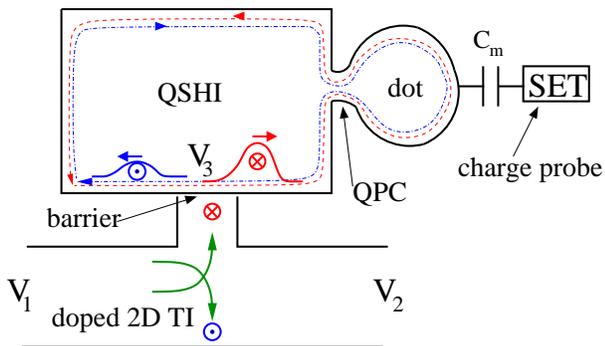}
\caption{Sketch of the first proposed device (not to scale).
A top gate drives the upper half of the device into the QSHI regime, with counter propagating helical edge states shown in dot-dashed (blue) and dashed (red) lines.
Dotted and crossed circles describe net spin densities along the growth direction (i.e. perpendicular to this page).
Another top gate drives the lower half of the device into the metallic spin-Hall phase.
A longitudinal electric field ($V_2\neq V_1$) in the doped 2D TI produces a spin accumulation at the tunnel barrier, which is injected into the QSHI edge states when $V_3-V_1=V_{\rm SD}\neq 0$.  
The spin densities of the two counterpropagating pulses emanating from the injection region are opposite in direction and unequal in magnitude (see Appendix).  
}
\label{fig:device}
\end{center}
\end{figure}

\begin{figure}[t]
\begin{center}
\includegraphics[scale=0.42]{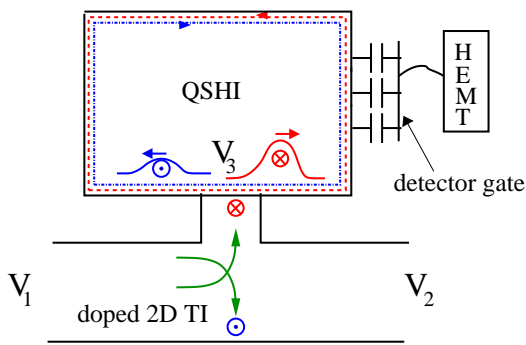}
\caption{Sketch of the second proposed device (not to scale). The injector is the same as in Fig.~\ref{fig:device}. The top gate that drives the upper half of the device into the QSHI phase is split in two pieces: one of them is a detector gate coupled to a HEMT (shown in the figure); the other gate covers the rest of the edge so as to avoid inhomogeneities in the Luttinger parameter.}
\label{fig:device2}
\end{center}
\end{figure}

The suggested devices are shown in Figs.~\ref{fig:device} and ~\ref{fig:device2}.
Fig.~\ref{fig:device} may be regarded as a topologically nontrivial counterpart to the quantum RC circuit discussed in ordinary 2D electron systems.~\cite{matveev1995,mora2010}
The upper part of the device is in the QSHI regime (chemical potential $\mu$ inside the bulk gap), and contains a narrow constriction (a quantum point contact or QPC) that opens onto a quantum dot.
A helical LL flows along the edges of the QSHI, as well as around the internal walls of the dot.  
The QPC is open, i.e. edge states outside the dot are seamlessly connnected with edge states inside the dot, and therefore there is no Coulomb blockade in the dot. 
The dot is coupled through a capacitance $C_m$ to a charge sensing circuit, such as an rf-single-electron-transistor (rf-SET),~\cite{field1993} which can detect small variations of charge inside the dot with little backaction.
The QPC is wider than the decay length of the edge states into the bulk,~\cite{zhou2008} yet narrower than the characteristic decay length of long-range Coulomb interactions.

The lower part of the device is a circuit that injects unidirectional electrons onto an edge of the QSHI.
Unidirectional injection can be achieved by either momentum-resolved tunneling from a parallel quantum wire~\cite{steinberg2008} or by tunneling from a magnetic scanning electron microscope,~\cite{das2011} both of which require magnetic elements or fields.
In Fig.~\ref{fig:device} we display an alternative mechanism, which is not only all-electric but also available in current experiments on 2D TIs.~\cite{brune2011}
Split gates drive the injector into a metallic spin-Hall phase ($\mu$ within a bulk band).
Owing to the spin-Hall effect,~\cite{she} a longitudinal electric field in the injector leads to a nonequilibrium spin accumulation at the tunnel barrier.
Accordingly, electrons injected onto the QSHI edge are momentum-resolved.

Fig.~\ref{fig:device2} is a quantum-spin Hall counterpart to the circular mesas employed for detection of edge magnetoplasmons in ordinary quantum Hall systems,\cite{ashoori1992} and thus we refer to it as a {\em TI mesa}.
The injector part of Fig.~\ref{fig:device2} is the same as in Fig.~\ref{fig:device}.
However, in lieu of a quantum dot, a finite segment of the 2D TI edge is coupled to an extended gate.
This gate detects charge passing underneath, and is connected to a high-electron-mobility-transistor (HEMT) by a conducting wire.
As elaborated below, an important difference between Fig.~\ref{fig:device2} and the setup of Ref.~[\onlinecite{ashoori1992}] is that the former requires additional gates in order to avoid plasmon backscattering. 
This difference originates from the fact that edge magnetoplasmons propagate along chiral edge states (where no backscattering is possible), whereas the fractional charges we study propagate along helical edge states (where backscattering is possible). 

\section{TI RC Circuit}

This section is directed to theoretically modeling the device of Fig.~\ref{fig:device}.
\subsection{Model}
\begin{figure}[t]
\begin{center}
\includegraphics[scale=0.4]{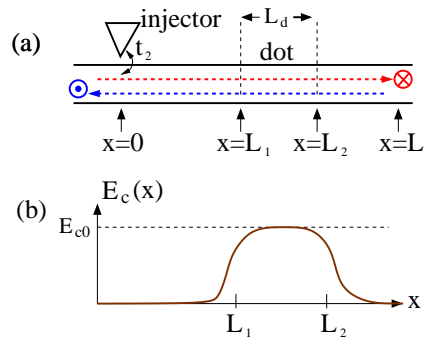}
\caption{(a) Unfolded line used for the model calculation. $x=0$ is the injection point, $x=L_1$ is the coordinate at the entrance to the quantum dot (in case of Fig.~\ref{fig:device}) or to the region beneath the detector gate (in case of Fig.~\ref{fig:device2}). 
 $x=L_2=L_1+L_d$ is the coordinate at the exit from the dot (in case of Fig.~\ref{fig:device}) or from the region beneath the detector gate (in case of Fig.~\ref{fig:device2}). 
$x=L$ coincides with $x=0$. 
(b) Spatial profile of the charging energy in Fig.~(\ref{fig:device}). The transition from $E_c(x)\simeq 0$ outside the dot to $E_c(x)\simeq E_{c0}$ inside the dot occurs smoothly.} 
\label{fig:toy}
\end{center}
\end{figure}
Away from commensurate fillings (so that Umklapp scattering may be neglected), the low-energy properties of the QSHI in Fig.~\ref{fig:device} can be modeled with a Hamiltonian  ${\cal H}={\cal H}_{\rm LL}+{\cal H}_{\rm bs}+{\cal H}_c+{\cal H}_T$ defined on an unfolded line with periodic boundary conditions (Fig.~\ref{fig:toy}a).

\begin{equation}
\label{eq:ll}
{\cal H}_{\rm LL}=(\hbar u/2)\int_0^L dx [(1/g)(\partial_x\phi)^2+g (\partial_x\theta)^2]
\end{equation}
is the unperturbed Hamiltonian of a helical LL~\cite{wu2006} (identical to that of a spinless LL) with a plasmon propagation velocity $u$. 
$\phi$ and $\theta$ are bosonic fields describing entangled spin and charge degrees of freedom and satisfying $[\theta(x),\partial_y\phi(y)]=i\delta(x-y)$. 
$L$ is the total perimeter of the QSHI region. 

\begin{equation}
{\cal H}_{\rm bs}=\alpha \hbar v_F\sum_\sigma\Psi_\sigma^\dagger(L_2)\Psi_\sigma(L_1)+{\rm h.c.}
\end{equation}
is the single-particle backscattering Hamiltonian, which encodes spin-conserving interedge tunneling at the QPC. 
$\alpha$ is the backscattering amplitude, $v_F=u g$ is the Fermi velocity, $\sigma=\uparrow,\downarrow$ is the spin direction, $\Psi_\sigma=\exp[i\sqrt{\pi}(\theta-\sigma\phi)]/\sqrt{2\pi a}$ represents a right-moving (with $\sigma=\downarrow$) or left-moving (with $\sigma=\uparrow$) free fermion at the edge, and $a$ is a short-distance cutoff.   
Below we justify ${\cal H}_{\rm bs}= 0$.

For a small enough quantum dot, the long-range part of the Coulomb interaction (not included in Eq.~\ref{eq:ll}) leads to a charging Hamiltonian of the form 
\begin{equation}
\label{eq:char}
{\cal H}_c\simeq E_{c0}\left[\left(\phi(L_1)-\phi(L_2)\right)/\sqrt{\pi}-N_g\right]^2.
\end{equation}
Eq.~(\ref{eq:char}) relies on a constant capacitance model, which may apply approximately in the quantum spin-Hall regime.
$E_{c0}$ is the charging energy of the dot, nonzero only for $x\in(L_1,L_2)$.
In HgTe quantum wells, we estimate $E_{c0}\simeq 0.1 {\rm meV}$  for a dot of perimeter $L_d\equiv(L_2-L_1)\gtrsim 1\mu{\rm m}$.
$\hat{Q}=\int_{L_1}^{L_2} dx \partial_x\phi/\sqrt{\pi}=[\phi(L_1)-\phi(L_2)]/\sqrt{\pi}$ is the edge charge of the dot (in units of $e$) and $N_g$ is the edge charge induced by a gate voltage applied to the dot.
The bulk charge of the QSHI is constant at energy scales of interest, and does not enter in our model.  

In absence of single-particle backscattering at the QPC, ${\cal H}_c$ enforces $\langle \hat{Q}(t)\rangle= N_g$ at temperatures below $E_{c0}$,~\cite{matveev1995} with $Q_{\rm rms}\equiv(\langle\hat{Q}(t)^2\rangle-N_g^2)^{1/2}\simeq [e^2 g^2 \ln(\mu/\xi)-N_g^2]^{1/2}\sim O(e)$.
Here $\xi\simeq E_{c0}$ when $E_{c0}\gg \hbar u/L_d$, and $\xi\simeq \hbar u/L_d$ when $E_{c0}\ll \hbar u/L_d$.
In TI quantum dots the ratio between $E_{c 0}$ and  $\hbar u/L_d$ is independent of $L_d$. 
In the case of HgTe quantum wells we expect $E_{c0}\simeq \hbar u/L_d$, whereas in InAs/GaSb/AlSb quantum wells~\cite{knez2011} we expect $E_{c0}\gg \hbar u/L_d$ due to significantly slower edge states.
By taming equilibrium charge fluctuations, the quantum dot facilitates a detection of injected (nonequilibrium) charges.

Finally,
\begin{equation}
{\cal H}_{T}=t_2 \Psi^\dagger_{2\uparrow}(0)\Psi_\uparrow(0)+{\rm h.c.}
\end{equation}
is the tunneling Hamiltonian describing unidirectional (i.e. spin-polarized) charge injection onto the helical liquid. 
$t_2$ is the tunneling amplitude and $\Psi_{2\uparrow}$ is a spin-up fermion operator in the injector. 
For simplicity we assume that injection occurs at a single point ($x=0$), which is appropriate when the linear dimensions of the tunnel barrier are smalller than $L_d$.

\subsection{Elastic Charge Backscattering}

A key advantage of the TI RC circuit in comparison to topologically trivial devices made of ordinary quantum wires and quantum dots is that it (ideally) eliminates both plasmon and single-particle elastic backscattering, thereby allowing for a real-time measurement of fractional charges.

On one hand, time-reversal symmetry bans elastic single-particle backscattering away from the QPC.
At the QPC, ${\cal H}_{\rm bs}$ can open a gap in the energy spectrum of the edge states.~\cite{zhou2008,teo2009}
If the QPC is wide compared to $\hbar v_F/\mu$, the fate of ${\cal H}_{\rm bs}$ can be analyzed perturbatively. 
We begin from its bosonized form 
\begin{equation}
\label{eq:h_bs}
{\cal H}_{\rm bs}= \tilde{\alpha}\cos[\sqrt{\pi}(\phi(L_2)-\phi(L_1))]\cos[\sqrt{\pi}(\theta(L_2)-\theta(L_1))]\nonumber,
\end{equation}
where $\tilde{\alpha}=2\alpha\hbar v_F/(\pi a)$. 
At temperatures $T<E_{c0}$, $\hat{Q} \simeq \langle\hat{Q}\rangle=N_g$ and thus ${\cal H }_{\rm bs}\simeq \tilde{\alpha} \cos(\pi N_g)\cos[\sqrt{\pi}(\theta(L_2)-\theta(L_1))]$.
From here we derive the renormalization group flow 
\begin{equation}
\frac{d\tilde{\alpha}}{d l}=\left(1-\frac{1}{2 g}\right) \tilde{\alpha},
\end{equation}
where $l=\ln(E_{c0}/T)$.
Hence, single-particle backscattering is relevant for $g>0.5$ and irrelevant for $g<0.5$.
Multi-particle interedge scattering processes have been ignored throughout because they are either irrelevant or reduced to c-numbers by the charging energy.
Likewise, the interedge electrostatic coupling at the QPC, $\beta\partial_x\phi(L_1)\partial_x\phi(L_2)$, is irrelevant ($d\beta/dl=-\beta$).
In HgTe quantum wells,~\cite{konig2007} various estimates\cite{teo2009,strom2010} yield $g\sim 0.5-0.9$.
For this range, $\alpha$ is relevant and grows large ($\sim 1$) as $T\to 0$.
Fortunately, the gate voltage can be chosen such that $N_g=1/2$, which eliminates ${\cal H}_{\rm bs}$ completely for any $g$.
Alternatively, $g<0.5$ might be achieved through gate engineering.~\cite{blanter1998} 

On the other hand, plasmon backscattering away from the QPC is suppressed assuming that $g$ is uniform along the edge.
In order to satisfy this assumption, the gate electrodes must be engineered in such a way that the capacitance per unit length between the edge and the nearest gate is uniform across the QSHI.
However, it is not necessary that the gate voltage $V_g$ be spatially uniform because $\int dx V_g(x) \partial_x\phi$ can be eliminated by a shift in $\phi$, without renormalizing $g$.   
At the QPC, the TI RC circuit prevents a mismatch in $g$ and is thus free from plasmon backscattering because (i) charge entering in a TI quantum dot flows along a helical LL at the inner wall of the dot, and (ii) the charging energy in the dot (which can be treated exactly) is found not to alter the value of $g$.
In contrast, an interface between a topologically trivial quantum wire and dot would inevitably lead to a mismatch in $g$ across the QPC. 
Indeed, ordinary semiconductor quantum dots are 2D Fermi liquids ($g=1$), whereas the quantum wire is a Luttinger liquid with $g<1$.\cite{caveat2}

\section{TI Mesa }

In this section we model the setup of Fig.~\ref{fig:device2}, which is inspired by the time-resolved detection of edge magnetoplasmons in ordinary quantum Hall systems.\cite{ashoori1992} 
Unlike in chiral quantum-Hall systems, where the Luttinger parameter is a topological invariant immune to external gates, coupling a segment of a 2D TI edge to a detector gate will result in a change of $g$ for that particular section of the wire.~\cite{blanter1998}
Since $v_F$ is independent of energy in the linear dispersion regime, a mismatch of $g$ would be tantamount to a mismatch in the plasmon velocity $u=v_F/g$.
This would induce plasmon backscattering\cite{lehur2008} and thus ruin the measurement of fractional charges.

In order to avoid this problem and keep $g$ uniform along the edge, we require that the rest of the edge be coupled to another gate located at the same distance from the QSHI edge as the detector gate.
Namely, once again the gate electrodes must be engineered in such a way that the capacitance per unit length between the edge and the nearest gate is uniform across the QSHI.  
With this proviso, the Hamiltonian for Fig.~\ref{fig:device2} can be written as
\begin{equation}
\label{eq:mesa}
{\cal H}={\cal H}_{LL}-V_g \hat{Q}+{\cal H}_T,
\end{equation}
where $\hat{Q}=[\phi(L_1)-\phi(L_2)]/\sqrt{\pi}$ is the edge charge contained along the segment covered by the detector gate ($x\in(L_1,L_2)$), and $V_g$ is the gate voltage therein. 
Even though Eq.~(\ref{eq:mesa}) contains no explicit electrostatic charging energy, a term $\sim (\hbar u/L_d) \hat{Q}^2$ emerges in the effective action for $(\phi(L_1)-\phi(L_2))$ after integrating out gapless modes, where $L_d=L_2-L_1$.
A simple calculation yields $Q_{\rm rms}\simeq [e^2 g^2 \ln(\mu L_d/\hbar u)-N_g^2]^{1/2}\sim O(e)$.
Therefore, the effective theories for a TI RC circuit with ${\cal H}_{\rm bs}=0$ and a TI mesa  are formally identical insofar as $g$ is spatially uniform.
 
In view of this similarity, and everything else being equal, the TI mesa  appears to be a more convenient platform than a TI RC circuit for the time-resolved measurement of fractional charges.
Perhaps a practical advantage of having a dot is that the fractional pulses spend longer time (by a factor $\pi$) under the detector gate than they would if the dot had been absent.

\section{Charge Fractionalization}

The aim of this section is to determine how the charge of the quantum dot (in case of Fig.~\ref{fig:device}) or the charge under the detector gate (in case of Fig.~\ref{fig:device2}) change when a weak bias voltage $V_{\rm SD}$ is applied between the injector and the QSHI.
To that end we compute $\delta Q(t)=\int_{-\infty}^t dt' I_{\rm net}(t')$, where $I_{\rm net}=I(L_1,t)-I(L_2,t)$, $I(x,t)=\langle\hat{I}(x,t)\rangle$ is the current at point $x$ along the edge, and $\hat{I}=(e v_F/\sqrt{\pi}) \partial_x\theta$ is the current operator for the helical liquid.
The outcome of the calculation is identical for the two proposed devices.

Conventional wisdom~\cite{lehur2008} dictates 
\begin{equation}
\label{eq:i1}
I(x,t)=\frac{1}{2}\sum_\eta\langle {\cal T}_K\hat{I}(x,t_\eta) e^{-\frac{i}{\hbar}\int_K dt'{\cal H}_T(t')}\rangle,
\end{equation}
where ${\cal T}_K$ is the time-ordering operator in the Keldysh contour $K$,  and $\eta=\pm$ denotes the upper/lower branch of $K$. 
The expectation value is taken with respect to the ground state of ${\cal H}-{\cal H}_T$, which 
importantly is quadratic in bosonic fields and can be diagonalized exactly. 
Recognizing that $E_c(x)\simeq 0$ close to the injection site, the influence of the charging energy in Fig.~\ref{fig:device} can be neglected for $x\notin(L_1,L_2)$. 

When $V_{\rm SD}={\rm const}$, Eq.~(\ref{eq:i1}) results in uniform and constant $I(x,t)$.~\cite{lehur2008} 
Therefore $I_{\rm net}=0$ and $\delta Q=0$, i.e. the dc bias considered in previous theoretical studies is not suitable for measuring fractional charges.
Herein we consider a short bias pulse, $V_{SD}(t)=\gamma\delta(t-t_0)$.
A calculation outlined in the Appendix yields 
\begin{equation}
\label{eq:i3}
I(|x|\gtrsim 0,t)= {\rm sgn}(x)[(1+g\, {\rm sgn}(x))/2] I_0(t-t_0-|x|/u),
\end{equation}
where $x>0$ $(x<0)$ denotes right (left) from $x=0$,
\begin{equation}
I_0(y)=\frac{a^{\nu-1}}{2\pi^2 u^{\nu+1}}\frac{e |t_2|^2}{\hbar^2}\sin\left(\frac{e\gamma}{\hbar}\right)\Theta(y) {\rm Im}\left[\frac{i/\nu}{(a/u-i y)^\nu}\right]\nonumber,
\end{equation}
$\nu\equiv(g+g^{-1})/2$, $a/u\simeq 0^+$ and $\Theta$ is the step function. 
$I_0(t-t_0)$, which is narrowly peaked at $t=t_0$, equals the total injected current and thus $Q_0=\int dt I_0(t-t_0)$ is the total injected charge.

For completeness we evaluate the bias-induced spin-density at a point $x$ and time $t$ (see Appendix):
\begin{equation}
\label{eq:s3}
{\bf S}(|x|\gtrsim 0,t)=\hat{z}\,\,{\rm sgn(x)}\frac{1+g\,{\rm sgn}(x)}{4 g}\frac{I_0(t-t_0-|x|/u)}{e\,u}
\end{equation}
The spin-densities of the counterpropagating eigen modes are opposite in direction and unequal in magnitude [ratio$=(1-g)/(1+g)$].
We verify that $\int_0^L dx \,{\bf S}(x,t)=\hat{z} Q_0/2$.
This indicates that although spin-up and spin-down single-particle states are not eigenstates of the helical LL, the total spin is conserved in electron fractionalization.
This is similar to the fact that although right- and left-moving single particle states are not eigenstates of a nonchiral LL, the total momentum is conserved in electron fractionalization.  

Eqs.~(\ref{eq:i3}) and ~(\ref{eq:s3}) generalize the results of Ref.~[\onlinecite{das2011}] to a time-dependent pulse, and confirm that charge and spin densities travel along the edge with velocity $u$ and without changing shape.

In the TI RC circuit, every injected charge packet flows unobstructed through the QPC insofar as the constriction at the QPC occurs smoothly over a lengthscale of several $\hbar v_F/\mu$.
This condition, and the suppression of ${\cal H}_{\rm bs}$ justified above, establish momentum conservation at the QPC. 
Momentum conservation, in conjunction with particle conservation, implies~\cite{lehur2008} that every charge pulse incident from the injector will be perfectly transmitted into the dot because, as explained above, there is no mismatch in $g$ at the QPC. 

\begin{figure}[t]
\begin{center}
\includegraphics[scale=0.44]{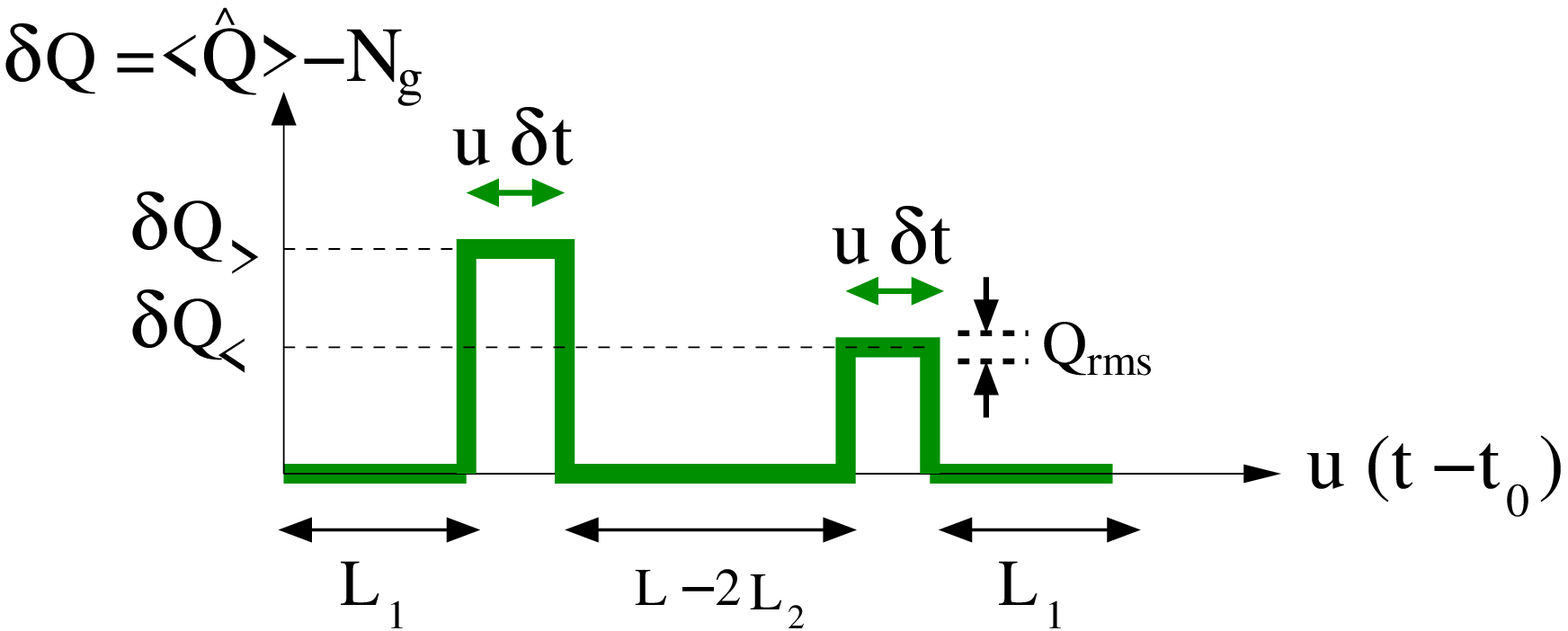}
\caption{$\delta Q (t)$ following a short pulse in the injector. $\delta t={\min}\{L_d/u,\hbar/E_{c0}\}$ in the device of Fig.~(\ref{fig:device}), and $\delta t=L_d/u$ in the device of Fig.~(\ref{fig:device2}). Only one time-period is shown. 
}
\label{fig:qfig}
\end{center}
\end{figure}

Eq.~(\ref{eq:i3}), the pulse-like character of $I_0$ and the perfect charge transmission at the QPC result in Fig.~\ref{fig:qfig}.
A fraction $(1+g)/2$ of $Q_0$ propagates along the edge with velocity $u$, enters the dot (or the detection region) at time $t=t_0+L_1/u$, exits it at time $t=t_0+L_1/u+\delta t$ and continues its counterclockwise trajectory around the device.
In a TI mesa , $\delta t$ equals the dwell time $t_{\rm dw}=L_d/u$. 
In a TI RC circuit, $\delta t={\rm min}\{t_{\rm dw},t_c\}$, where $t_c=\hbar/E_{c0}$ is the charging time. 
On the other hand, a fraction $(1-g)/2$ of $Q_0$ enters the dot at time $t=t_0+(L-L_2)/u$ and exits it at time $t=t_0+(L-L_2)/u+\delta t$, after which it continues its clockwise trajectory around the device.
At $t=t_0+L/u+\delta t-t_{\rm dw}$ the two pulses coincide at the injection point and the cycle repeats.
On each cycle the counterpropagating pulses meet at $x=0$ and $x=L/2$. 
At $x=L/2$ the pulses cannot recombine because they are orthogonal eigenstates of the helical liquid.
At $x=0$ they can in principle recombine (due to ${\cal H}_T$) and tunnel back to the injector; nevertheless charge conservation will then impose another pulse from the injector onto the helical edge state.  
$\delta Q$ is thus periodic in time, with two unequal peaks within each period: $\delta Q_>=(1+g) Q_0/2$ and $\delta Q_<=(1-g) Q_0/2$.
$\delta Q_</\delta Q_>=(1-g)/(1+g)$ gives the ratio of fractional charges, and the time period $L/u+\delta t-t_{\rm dw}\simeq L/u$ determines their speed.

Thus far we have assumed unidirectional (or fully spin polarized) injection of electrons, which can be realized by momentum-resolved tunneling.
However, the typical spin polarization for all-electrical injection is modest ($\lesssim 10\%$ with ballistic sources~\cite{nomura2005}). 
This can be accounted for by a straightforward generalization of Eq.~(\ref{eq:i3}), resulting in
\begin{equation}
\label{eq:ratio}
\delta Q_</\delta Q_>=(1- g P)/(1+ g P),
\end{equation}
where $P=(Q_{0\uparrow}-Q_{0\downarrow)}/Q_0$ is the spin polarization of the injected pulse.
For the setup of Fig.~\ref{fig:device}, $P=\lambda e E/(\hbar v_F n)$, where $E$ is the longitudinal electric field in the injector, $n$ is the carrier density in the doped 2D TI, and $\lambda\sim O(1)$ is a dimensionless measure of the strength of spin-orbit interactions.
Since $P$ and $g$ can be measured independently from one another (the latter through d.c. tunneling conductance measurements, the former by electrical~\cite{brune2011} or optical~\cite{nomura2005} means), Eq.~(\ref{eq:ratio}) provides a test for charge fractionalization.  

Even though Eq.~(\ref{eq:ratio}) has been derived for a pulse whose time width $\Delta t_{\rm pulse}$ is shorter than $\delta t$, it can be shown that Eq.~(\ref{eq:ratio}) is quantitatively applicable to wavelets with $\Delta t_{\rm pulse}\gg t_{\rm dw}$ as well, provided that $\delta Q_>$ and $\delta Q_<$ are understood as time-averaged charges over an interval $\gtrsim\Delta t_{\rm pulse}$.
Defining $t_{\rm tr}\equiv{\rm min}\{L-2 L_2,2 L_1\}/u$ as the shortest time interval in which $\delta Q=0$, $\Delta t_{\rm pulse}\ll t_{\rm tr}$ ensures that the two counterpropagating charge pulses will not overlap inside the dot. 

In order to time-resolve the fractional charge pulses, the rf-SET of Fig.~\ref{fig:device} and the HEMT of Fig.~\ref{fig:device2} must have frequency bandwidths $\Delta f> (2\pi t_{\rm tr})^{-1}$, which sets a lower bound on the size of the device. 
For an rf-SET, an optimistic assessment is $\Delta f\simeq 0.1-1\, {\rm GHz}$.~\cite{set} 
Then, in HgTe devices with $u\simeq 10^6 {\rm m/s}$, the required total edge length for the TI RC circuit is $L\gtrsim 1 {\rm mm}$.
For $L\gtrsim 1 {\rm mm}$, $L_1\simeq 500\mu{\rm m}$ and $L_d\simeq 5\mu{\rm m}$, it follows that $t_{\rm dw}\simeq t_c\lesssim 0.01 {\rm ns}$ and $t_{\rm tr}\lesssim 1 {\rm ns}$.
Moreover, when $\Delta t\simeq 0.1 {\rm ns}$, a good signal-to-noise ratio requires $Q_0\gtrsim 10^3 e$, so that the maxima of $\delta Q (t)$ exceed $Q_{\rm rms}$.  
A pulse that injects $1000$ electrons is still a relatively small perturbation compared to the typical number of electrons contained in the edge states ($\sim L\mu/\pi \hbar v_F\gtrsim 5\times 10^3$). 

Regarding the TI mesa  of Fig.~\ref{fig:device2}, typical HEMTs possess very high frequency bandwidths ($\Delta f\gtrsim 10 {\rm GHz}$) and are able to detect charge pulses at sub-nanosecond timescales.
Yet, this enhanced time resolution comes at a cost of a much reduced charge sensitivity.
For instance, in Ref.~[\onlinecite{ashoori1992}] the charge resolution was $\sim$ 100 electrons. 
In order to compensate for this a larger number of electrons must be injected, which may then lead to undesired inelastic and nonlinear effects.

\section{Discussion}


\subsection{Inelastic charge backscattering}

The experiments of Figs.~\ref{fig:device} and ~\ref{fig:device2} can be successfully implemented only if the time $t_{\rm in}$ for inelastic backscattering (which is permitted by time-reversal symmetry) is longer than $L/u$.
While this condition appears achievable at sufficiently low temperatures, current HgTe TI edges are plagued with quasi-2D conducting puddles~\cite{roth2009} that render $u t_{\rm in}\simeq 10 \mu{\rm m}\ll L$. 
It is likely that, as experiments improve, conducting droplets will become far less common. 
Alternatively, InAs/GaSb/AlSb quantum wells~\cite{knez2011} host 2D TIs with slow edge states ($v_F\simeq 3\times 10^4 {\rm m/s}$). 
Therefore, TI RC circuits fabricated with these materials could be small ($L\simeq 10\mu{\rm m}$) and still satisfy both $u t_{\rm in}> L$ and $\Delta f\gtrsim (2\pi t_{\rm tr})^{-1}$.  

\subsection{Influence of Rashba spin-orbit interaction}

Throughout this paper we have assumed QSHI edges whose spin quantization axes coincide with the growth direction of the quantum well.
This assumption no longer holds in presence of structural inversion asymmetry and its accompanying Rashba spin-orbit interaction (SOI), which can be sizeable in HgTe quantum wells.
In its simplest realization,~\cite{strom2010} Rashba SOI rotates the spin quantization axis of a 2D TI edge.
The rotation angle can be tuned by a gate acting on the QSHI, and the rotation axis varies from one edge of the device to another.
Consequently, the spin density that tunnels from the injector is no longer parallel or antiparallel to the spin quantization axis at the QSHI edge.

Let us suppose that the quantization axis of the edge states at the injection site is $\hat{n}\neq\hat{z}$. 
Then, Eq.~(\ref{eq:ratio}) remains valid if we redefine 
\begin{equation}
P=(Q_{0,+}-Q_{0,-})/Q_0,
\end{equation}
where $Q_{0,\pm}=Q_{0\uparrow}|\langle\uparrow|\hat{n},\pm\rangle|^2+Q_{0\downarrow}|\langle\downarrow|\hat{n},\pm\rangle|^2$ and  $|\hat{n},\pm\rangle$ are eigenstates of ${\boldsymbol\sigma}\cdot\hat{n}$.
Thus, even when the injected pulse is fully spin polarized ($Q_{0\uparrow}=Q_0$), $P<1$ due to $\hat{n}\neq\hat{z}$.
The fingerprints of Rashba SOI in charge fractionalization can be singled out by measuring the dependence of $P$ on the gate voltage applied to the QSHI.

Although Rashba SOI entails different spin quantization axes along different TI edges, there is no any elastic bacskcattering at the corners (assumed to be smooth on lengthscales $\sim\hbar v_F/\mu$). 
Instead, the edge states (which are now eigenstates of the full Hamiltonian including Rashba SOI) have spatially inhomogeneous spin textures.
The spin densities attached to the fractional charges precess as the pulses travel along the circuit (see Appendix).

\section{Conclusions}

We have theoretically designed two topological insulator circuits that may enable a hitherto elusive time-resolved measurement of fractional charges in nonchiral Luttinger liquids. 
The main advantage of these TI circuits in comparison with previous proposals (such as the three terminal quantum wire geometries) is that they can be made robust against elastic backscattering, and can be integrated with noninvasive charge sensors.
In a broader context, our work illustrates a concrete example of how TI devices may qualitatively outperform ordinary semiconductor devices.

Future research avenues include designing probes of quantum entanglement and calculating higher moments of charge (e.g. $\langle Q_i Q_j\rangle-\langle Q_i\rangle\langle Q_j\rangle$) in TI RC circuits and TI mesas containing multiple charge sensors. 

This project has benefited from a winter conference at the Aspen Center for Physics, and has been funded by a fellowship from Yale University (IG), by DOE Grant No. DE-FG02-08ER46541 (KLH), and by NSF Grant No. DMR-0803200 (KLH). 
IG acknowledges helpful comments from L.I. Glazman, K.A. Matveev and T.L. Schmidt. 
KLH acknowledges useful discussions with B. Halperin.

\appendix
\begin{widetext}
\section{Charge and spin fractionalization for a short injection pulse}
In this Appendix we provide calculational details leading to Eq. (2) of the main text.
We also calculate the spin polarization of the counterpropagating pulses, in order to justify the dotted and crossed circles ascribed to them in Fig. 1 of the main text.
We apply the nonequilibrium Keldysh formalism as presented for instance by A. Cr\'epieux {\em et al.} and by K. Le Hur {\em et al.} (Refs.~[\onlinecite{safi1995}] and ~[\onlinecite{lehur2008}], respectively).
However, details of our calculation are quite different from those references because we take a time-dependent bias.
Alternatively we could have employed ordinary linear and quadratic response functions; however, the advantage of using the Keldysh method lies in its time-ordered correlation functions, which lend themselves to convenient bosonization tricks. 
For simplicity we assume that the injected charge is 100\% spin-polarized along the z-axis, which is perpendicular to the page in Figs.~\ref{fig:device} and \ref{fig:device2}.
The generalization to partial spin polarization is straightforward, and it produces Eq.(\ref{eq:ratio}) of the main text. 
In addition, we note that the injected charge in the all-electrical devices of Figs. \ref{fig:device} and \ref{fig:device2} has no net magnetization in the XY plane.

\subsection{Total injected current}
The current injected at $x=0$ is given by
\begin{eqnarray}
\label{eq:i0}
I_0(t)&=&\frac{1}{2}\sum_\eta\langle{\cal T}_K \frac{i e}{\hbar}[{\cal H}_T(t_\eta),\Psi^\dagger_\uparrow(t_\eta)\Psi_\uparrow(t_\eta)] e^{-\frac{i}{\hbar}\int_K dt'{\cal H}_T(t')}\rangle\nonumber\\
&\simeq& -\frac{i e |t_2|^2}{\hbar^2}\sum_\eta\eta\int_{-\infty}^{\infty} dt'\sin\left(\int_{t'}^{t} V(\tilde{t}) d\tilde{t}\right) G^{\eta.-\eta}(t-t') G_2^{\eta,-\eta}(t-t'),
\end{eqnarray}
where $\eta=\pm$ denotes the forward/backward branch of the Keldysh contour, $V=(e/\hbar) V_{\rm SD}$ is the external bias (incorporated through a Peierls substitution), $G_2^{\eta,-\eta}(\tau=t-t')\equiv\langle{\cal T}_K\Psi_{2\uparrow}(0,t_\eta)\Psi_{2\uparrow}^\dagger(0,t'_{-\eta})\rangle=(2\pi u)^{-1}(a/u-i\eta \tau)^{-1}$ is a fermionic Green's function for the injector at $x=0$ and $G^{\eta,-\eta}(\tau)\equiv\langle{\cal T}_K\Psi^\dagger_\uparrow(0,t_\eta)\Psi_\uparrow(0,t'_{-\eta})\rangle= a^{\nu-1}(2\pi u^\nu)^{-1}(a/u-i \eta \tau)^{-\nu}$ is a fermionic Green's function for the helical LL at $x=0$.
We have assumed without loss of generality that the injector can be characterized as a free fermion 1D system.
In order to derive the second line of Eq.~(\ref{eq:i0}) we have followed the same steps as in Ref.~[\onlinecite{lehur2008}].
For $V(t)=(e/\hbar)\gamma\delta(t-t_0)$, we obtain
\begin{equation}
\label{eq:i02}
I_0(t)=\frac{a^{\nu-1}}{4\pi^2 u^{\nu+1}}\frac{2 e |t_2|^2}{\hbar^2}\sin\left(\frac{e}{\hbar}\gamma\right){\rm Im}\left[\frac{i/\nu}{[a/u-i(t-t_0)]^\nu}\right],
\end{equation}
in which derivation we have used $\sin\left(\int_{t'}^t V d\tilde{t}\right)=\sin(e\gamma/\hbar)[\Theta(t-t_0)\Theta(t_0-t')-\Theta(t'-t_0)\Theta(t_0-t)]$ and have performed the integration over $t'$.
Because $a/u=0^+$, $I_0(t)$ vanishes for all times except $t\to t_0$.

\subsection{Current along the helical LL}
The current flowing at a point $x$ of the helical liquid at time $t$ is given by
\begin{eqnarray}
I(x,t)&=&\frac{1}{2}\sum_\eta\langle{\cal T}_K \hat{I}(x,t_\eta)e^{-\frac{i}{\hbar}\int_K dt'{\cal H}_T(t')}\rangle\simeq\frac{1}{4\hbar^2}\sum_{\eta\eta_1}\langle{\cal T}_K \frac{e v_F}{\sqrt{\pi}}\partial_x\theta(x,t_\eta)\int_{-\infty}^\infty dt' dt'' {\cal H}_T(t'_{\eta_1}){\cal H}_T(t''_{-\eta_1})\rangle.
\end{eqnarray}
Following the same steps as in Ref.~[\onlinecite{lehur2008}], we reach
\begin{eqnarray}
I(x,t)&=& \frac{e v_F |t_2|^2}{2 \hbar^2}\sum_{\eta\eta_1}\int_{-\infty}^\infty dt'\int_{-\infty}^\infty dt''\sin\left(\int_{t''}^{t'} V d\tilde{t}\right) G^{\eta_1,-\eta_1}(t'-t'') G_2^{\eta_1,-\eta_1}(t'-t'')\nonumber\\
&&\times\partial_x[G_{\theta\theta}^{\eta,\eta_1}(x,t-t')+G_{\theta\phi}^{\eta,\eta_1}(x,t-t')-G_{\theta\theta}^{\eta,-\eta_1}(x,t-t'')-G_{\theta\phi}^{\eta,-\eta_1}(x,t-t'')],
\end{eqnarray} 
where $G_{\theta\theta}$ and $G_{\theta\phi}$ are the Green's functions of the bosonic fields (for a succint summary see e.g. A. Cr\'epieux {\em et al.}). 
For $V(t)=(e/\hbar)\gamma\delta(t-t_0)$,  $\sin\left(\int_{t'}^{t''} V d\tilde{t}\right)=\sin(e\gamma/\hbar)[\Theta(t'-t_0)\Theta(t_0-t'')-\Theta(t''-t_0)\Theta(t_0-t')]$ and after a little algebra we get
\begin{eqnarray}
I(x,t)&=&\frac{e v_F |t_2|^2}{2 \hbar^2}\sin\left(\frac{e}{\hbar}\gamma\right)\sum_{\eta\eta_1}\nonumber\\
&&\times\Big[\int_{t_0}^{\infty} dt'\int_{-\infty}^{t_0} dt'' \left(f^{\eta,\eta_1}(x,t-t') g^{\eta_1,-\eta_1}(t'-t'')+f^{\eta,-\eta_1}(x,t-t') g^{\eta_1,-\eta_1}(t''-t')\right)\nonumber\\
~~~~~&&-\int_{-\infty}^{t_0} dt'\int_{t_0}^{\infty} dt''\left(f^{\eta,\eta_1}(x,t-t') g^{\eta_1,-\eta_1}(t'-t'')+f^{\eta,-\eta_1}(x,t-t') g^{\eta_1,-\eta_1}(t''-t')\right)\Big],
\end{eqnarray}
where $f^{\eta,\eta_1}(x,\tau)\equiv\partial_x[G^{\eta,\eta_1}_{\theta\theta}(x,\tau)+G^{\eta,\eta_1}_{\theta\phi}(x,\tau)]$ and $g^{\eta,-\eta_1}(\tau)\equiv G^{\eta_1,-\eta_1}(\tau) G_2^{\eta_1,-\eta_1}(\tau)$.
Doing the $t''$ integral and rearranging terms we arrive at
\begin{equation}
I(x,t)=\frac{a^{\nu-1}}{4\pi^2 u^{\nu+1}}\frac{e v_F |t_2|^2}{\hbar^2}\sin\left(\frac{e}{\hbar}\gamma\right)
\sum_{\eta\eta_1}\int_{-\infty}^\infty dt' f^{\eta,\eta_1}(x,t-t')\frac{i}{\nu}\eta_1\frac{1}{[a/u-i\eta_1(t'-t_0)]^\nu}.
\end{equation}
Now we note
\begin{eqnarray}
&&\sum_{\eta\eta_1}\int_{-\infty}^\infty dt' f^{\eta,\eta_1}\eta_1\frac{i/\nu}{[a/u-i\eta_1(t'-t_0)]^\nu}=
\int_{-\infty}^\infty dt'\left[(f^{++}+f^{-+})\frac{i/\nu}{[a/u-i(t'-t_0)]^\nu}-(f^{--}+f^{+-})\frac{i/\nu}{[a/u+i(t'-t_0)]^\nu}\right].\nonumber
\end{eqnarray}
It will be clear below that the $f$-functions can be written as $PV(1/x)-i\pi\delta(x)$, where $PV$ is the principal value, $\delta(x)$ is the Dirac delta and $x$ is a real number (or parameter). 
After verifying that the principal value contributions vanish upon integrating over $t'$ from $-\infty$ to $+\infty$, and that $I(x,t)$ is a real number (as it should), we write
\begin{eqnarray}
I(x,t) &=&-\frac{a^{\nu-1}}{4\pi^2 u^{\nu+1}}\frac{e v_F |t_2|^2}{\hbar^2}\sin\left(\frac{e}{\hbar}\gamma\right)\int_{-\infty}^{\infty} dt' {\rm Im}(f^{++}+f^{-+}-f^{--}-f^{+-}){\rm Im}\left[\frac{i/\nu}{[a/u-i(t'-t_0)]^\nu}\right],
\end{eqnarray}
where the argument of the $f$-functions is $(x,t-t')$.
Borrowing the standard expressions for $G^{\eta\eta'}_{\theta\theta}$ and $G^{\eta\eta'}_{\theta\phi}$, we obtain
\begin{eqnarray}
\label{eq:f}
&&f^{++}(x,\tau)+f^{-+}(x,\tau)-f^{--}(x,\tau)-f^{+-}(x,\tau)\nonumber\\
&=&\Theta(\tau)\left[\frac{1-g}{2\pi}\left(\frac{i}{a+i v_F\tau+i g x}+\frac{i}{a-i v_F\tau-i g x}\right)-\frac{1+g}{2\pi}\left(\frac{i}{a-i v_F\tau+i g x}+\frac{i}{a+i v_F \tau-i g x}\right)\right],
\end{eqnarray}
where $v_F=u g$ and $a$ can be regarded as $0^+$.
The bosonic Green's functions used for the derivation of Eq.~(\ref{eq:f}) are Cr\'epieux's expressions multiplied by a factor of $2$, because we are dealing with spinless fermions [$\Psi\sim\exp(i\sqrt{\pi}(\theta\pm\phi)$].
Besides, Eq.~(\ref{eq:f}) neglects the charging energy, and is therefore valid provided that $x\notin(L_1,L_2)$.

Hence,
\begin{equation}
{\rm Im}(f^{++}+f^{-+}-f^{--}-f^{+-})=\left[(1-g)\delta(v_F\tau+g x)-(1+g)\delta(v_F\tau-g x)\right]\Theta(t-t'),
\end{equation}
leading us to
\begin{eqnarray}
I(x,t)&=&-\frac{a^{\nu-1}}{4\pi^2 u^{\nu+1}}\frac{e v_F |t_2|^2}{\hbar^2}\sin\left(\frac{e}{\hbar}\gamma\right)\int_{-\infty}^t dt' [(1-g)\delta(v_F (t-t')+g x)-(1+g)\delta(v_F(t-t')-g x)]{\rm Im}\left[\frac{i/\nu}{[a/u-i (t'-t_0)]^\nu}\right]\nonumber\\
&=&\frac{a^{\nu-1}}{4\pi^2 u^{\nu+1}}\frac{e |t_2|^2}{\hbar^2}\sin\left(\frac{e}{\hbar}\gamma\right)(1+g){\rm Im}\left[\frac{i/\nu}{[a/u-i(t-t_0-x/u)]^\nu}\right]\Theta(x)\nonumber\\
&&-\frac{a^{\nu-1}}{4\pi^2 u^{\nu+1}}\frac{e |t_2|^2}{\hbar^2}\sin\left(\frac{e}{\hbar}\gamma\right)(1-g){\rm Im}\left[\frac{i/\nu}{[a/u-i(t-t_0+x/u)]^\nu}\right]\Theta(-x)\nonumber\\
&=&\frac{1+g}{2}I_0(t-t_0-x/u)\Theta(x)-\frac{1-g}{2}I_0(t-t_0+x/u)\Theta(-x),
\end{eqnarray}
which is Eq. (2) of the main text and satisfies current conservation.

\subsection{Spin polarization of the fractional charges}

The bias-induced spin-density at a point $x$ of the helical liquid at time $t$ is given by
\begin{equation}
\label{eq:spin}
{\bf S}(x,t)=\frac{1}{2}\sum_\eta\langle{\cal T}_K \hat{{\bf S}}(x,t_\eta)e^{-\frac{i}{\hbar}\int_K dt'{\cal H}_T(t')}\rangle,
\end{equation}
where $\hat{\bf S}=(1/2)\sum_{\alpha\beta}\Psi^\dagger_\alpha {\boldsymbol\sigma}_{\alpha\beta}\Psi_\beta$ ($\alpha,\beta\in\{\uparrow,\downarrow\}$).  
The bosonization of $S^z$ immediately yields $S^z(x,t)=I(x,t)/(2 e v_F)$.
Likewise, a quick inspection suffices to show that $S^x(x,t)=S^y(x,t)=0$.
In effect, a perturbative expansion of (say) $S^x$ in $t_2$ produces expectation values of strings fermion creation and annihilation operators, in which the number of right-movers (spin-down) and left-movers (spin-up) differs by one. 
It is easy to verify that those expectation values vanish (see e.g. Appendix C of Giamarchi in Ref.~[\onlinecite{deshpande2010}]).
In sum, the nonequilibrium spin polarization in the QSHI of Fig. 1 is
\begin{equation}
{\bf S}(x,t)=\hat{z}\left[\frac{1+g}{4 g}\frac{I_0(t-t_0-x/u)}{e\,\, u}\Theta(x)-\frac{1-g}{4 g}\frac{I_0(t-t_0+x/u)}{e\,\,u}\Theta(-x)\right].
\end{equation}
The spin-densities of the counterpropagating pulses are opposite in direction and unequal in magnitude [ratio$=(1-g)/(1+g)$].

It is worth noting that $S^x$ and $S^y$ would not have vanished if the injected spin polarization was not parallel to the  spin quantization axis $\hat{z}$ of the edge states.
Such is the case e.g. in presence of Rashba spin-orbit interactions.
In this situation, $S^x$ and $S^y$ show oscillatory behavior characteristic of spin precession; for example,
\begin{equation}
S^x(x>0,t)\propto \cos(2 k_F x) \delta(x-u (t-t_0))\,\,\,\mbox{   and   }\,\,\, S^y(x>0,t)\propto \sin(2 k_F x) \delta(x-u (t-t_0)),
\end{equation}
where we have used $\langle\psi_\uparrow^\dagger (x,t)\psi_\downarrow (x,t)\rangle\propto \exp(2 i k_F x)$ and have assumed that the injected charge is partly spin polarizated along $\hat{x}$.
The precession rate is independent of the strength of electron-electron interactions.

\end{widetext}


\begin{thebibliography}{100}
\bibitem{caveat2} Strictly speaking there is a way to avoid plasmon backscattering in a conventional 2D dot connected to a 1D wire, in spite of the inevitable discontinuity in $g$. The trick consists of tuning the density of electrons in the 2D dot in such a way that the Fermi velocity in the dot matches the plasmon propagation velocity in the 1D wire. However, this requires substantial fine tuning.
\end{thebibliography}
\end{document}